\documentclass[aps,twocolumn,pra,superscriptaddress,groupaddress,showpacs,preprintnumbers,amsmath,amssymb]{revtex4}
\newcommand{\beq}{\begin{equation}}
\newcommand{\eeq}{\end{equation}}
\newcommand{\beqnar}{\begin{eqnarray}}
\newcommand{\eeqnar}{\end{eqnarray}}
\newcommand{\tup}{\textup}
\newcommand{\Tr}{\operatorname{Tr}}

\newcommand{\K}{{\bf k}}
\newcommand{\Pmom}{{\bf p}}

\newcommand{\sigbf}{{\boldsymbol \sigma}}

\newcommand{\upar}{\uparrow}
\newcommand{\dnar}{\downarrow}
\newcommand{\sgn}{\textup{sgn}}
\ifx\pdfoutput\undefined
\usepackage[dvips]{graphicx}
\usepackage[dvips]{color}
\usepackage{epsfig}
\else
\usepackage[pdftex]{graphicx}
\usepackage{epstopdf}
\fi
\usepackage{dcolumn}
\usepackage{bm}
\begin{document}
\title{Non-Abelian topological order in non-centrosymmetric superconductors with broken time reversal symmetry}
\author{Parag Ghosh}
\affiliation{Department of Physics and Astronomy, George Mason University, Fairfax, Virginia, 22030, USA and \\ 100 Bureau Drive, Stop 8410, NIST, Gaithersburg, Maryland, 20899-8410, USA}
\author{Jay D. Sau}
\affiliation{Condensed Matter Theory Center and Joint Quantum Institute, Department of Physics, University of Maryland, College Park, Maryland, 20742-4111, USA}
\author{Sumanta Tewari}
\affiliation{Department of Physics and Astronomy, Clemson University, Clemson, South Carolina, 29634, USA}
\affiliation{Condensed Matter Theory Center and Joint Quantum Institute, Department of Physics, University of Maryland, College Park, Maryland, 20742-4111, USA}
\author{S. Das Sarma}
\affiliation{Condensed Matter Theory Center and Joint Quantum Institute, Department of Physics, University of Maryland, College Park, Maryland, 20742-4111, USA}
\date{\today}
\begin{abstract}
We consider two-dimensional non-centrosymmetric superconductors, in which the order parameter is a mixture of $s$-wave and $p$-wave parts, in the presence of an externally induced Zeeman splitting. We derive the conditions under which the system is in a non-Abelian phase. By considering the non-degenerate zero-energy Majorana solutions of the Bogoliubov-de Gennes (BdG) equations for a vortex and by constructing a topological invariant, we show that the condition for the non-Abelian phase to exist is completely independent of the triplet pairing amplitude. The existence condition for the non-Abelian phase derived from the real space solutions of the BdG equations involves the Pfaffian of the BdG Hamiltonian at $k=0$, which is completely insensitive to the magnitude of the $p$-wave component of the order parameter. We arrive at the same conclusion by using the appropriate topological invariant for this case. This is in striking contrast to the analogous condition for the time-reversal \emph{invariant} topological phases, in which the amplitude of the $p$-wave component must be larger than the amplitude of the $s$-wave piece of the order parameter. As a by-product, we establish the intrinsic connection between the Pfaffian of the BdG Hamiltonian at $k=0$ (which arises at the BdG approach) and the relevant $\mathbb{Z}$ topological invariant.
\end{abstract}
\pacs{03.67.Lx, 71.10.Pm, 74.45.+c}
\maketitle
In quantum mechanics, statistics is defined as the transformation rule for a many-particle wave function under a pair-wise interchange of the particle coordinates. Bosons and fermions are the simplest of the quantum particles in that the corresponding many-body wave functions undergo a trivial transformation -- multiplication by 1 or -1, respectively, -- under an interchange of the quantum numbers of any two particles. In the special case of $(2+1)$ dimensions, in which simple permutation of the particle coordinates and actual exchange operations can be inequivalent, the statistics of particles can be more complex.~\cite{leinaas_nuovo, wilczek1, wilczek_book} Anyons are those quantum particles
for which the two-dimensional many body wave-function, under pair-wise exchange of the particle coordinates,
 receives a phase factor $e^{i\theta}$, where the statistical angle $\theta$ can take \emph{any} value between $0$ (bosons) and $\pi$ (fermions). An example of anyons are the quasiparticle excitations of the celebrated $\nu=\frac{1}{3}$ fractional quantum Hall state in which the quasiparticles have anyonic statistics with the statistical angle $\theta=\frac{\pi}{3}$.~\cite{Laughlin_PRL_1983}

In even more complex two-dimensional systems, if the many-body ground state wave function happens to be a linear combination of states from a degenerate subspace,
a pair-wise exchange of the particles can unitarily \emph{rotate} the wave function of the ground state in this subspace.
Consequently, the statistics is non-Abelian,~\cite{Kitaev, nayak_RevModPhys'08} and the corresponding system is a non-Abelian quantum system.
Non-Abelian quantum systems in the so-called Ising topological class, \cite{nayak_RevModPhys'08}
    are characterized by quasiparticle excitations called Majorana fermions. In superconducting systems, which are our focus in this paper, Majorana fermions arise as non-degenerate, spatially localized, zero-energy quasiparticle excitations bound to defects (such as vortices and boundaries) of the superconducting order parameter. Because of the non-degeneracy, the Majorana fermion solutions of the excitation spectrum are topologically protected, \emph{i.e.,} any local perturbation to the BdG Hamiltonian near an order parameter defect cannot move a Majorana fermion solution away from zero energy. The second quantized operators, $\gamma_i$, corresponding to these zero energy excitations are self-hermitian, $\gamma_i^{\dagger}=\gamma_i$, which is in striking contrast to ordinary fermionic (or bosonic) operators for which $c_i \neq c_i^{\dagger}$. The Majorana fermions, which are actually more like half-fermions, were envisioned \cite{Majorana} by E. Majorana in 1935 as fundamental constituents of nature (\emph{e.g.} neutrinos are thought to be Majorana, rather than Dirac, fermions).  Majorana modes are intriguing \cite{Wilczek-3} because each Majorana particle is its own anti-particle unlike ordinary fermions where electrons and positrons (or holes) are distinct.

There has been a growing interest in realizing Majorana fermions in the laboratory for topological quantum information processing purposes.~\cite{nayak_RevModPhys'08} Majorana fermions  can be used to build topologically protected quantum
memory applications. Even more importantly, they can be used in topological quantum computation (TQC) along with supplementary unprotected
quantum gates requiring only small amounts of error corrections.~\cite{Bravyi} TQC, in contrast
   to ordinary quantum computation, would not require
   any quantum error correction since the Majorana
   excitations are immune to local noise by virtue of
   their non-local topological nature.~\cite{Stern, nayak_RevModPhys'08}  One of the first candidate systems supporting Majorana fermion excitations that have been proposed are the non-Abelian $\nu=5/2$ fractional quantum Hall state \cite{Read_Green} and chiral $p$-wave superconductors \cite{Ivanov_PRL_2001} or superfluids. Recently, Fu and Kane \cite{FuKane_proposal_PRL_2008} proposed that the surface of a strong topological insulator (STI) in proximity to an $s$-wave superconductor can support a single non-degenerate Majorana mode bound to the vortex cores. Later, Sau {\it et al.} \cite{sau_prl_2010} suggested that the Fu and Kane set up can be significantly simplified by replacing the STI with an ordinary semiconductor with strong spin-orbit coupling. It was shown that in the presence of a proximity induced pure $s$-wave pairing potential $\Delta_s$ and a Zeeman splitting, the system under certain conditions \cite{sau_prl_2010, alicea} can support a non-degenerate Majorana fermion in the vortex core. Following this, it was quickly realized \cite{unpublished} that the one-dimensional version of the same set up, a semiconducting Majorana nanowire with zero-energy states at the two ends, would likely be an easier system to explore the physics of Majorana fermions, since the gap to the higher-energy non-topological excitations (so-called mini-gap) is of order $\Delta_s$ (there are no other sub-gap states other than the Majorana states). Further, the Zeeman splitting in this case can be introduced by a magnetic field parallel to the 
   superconductor,~\cite{lutchyn, gil} making a proximate magnetic insulator~\cite{sau_prl_2010} unnecessary. A comprehensive discussion of the spin-orbit coupled semiconductor with proximity induced $s$-wave superconductivity has been given in Ref.~[\onlinecite{long-paper}].

Even though spin-orbit coupled systems with proximity induced $s$-wave superconductivity have some important advantages, such as
  mini-gap $\sim \Delta_s$ much larger than the usual mini-gap $\sim \frac{\Delta_t^2}{E_F}$ ($\Delta_t$ is the $p$-wave gap and $E_F$ is the Fermi energy) in chiral-$p$-wave superconductors, a crucial new requirement is good superconducting proximity effect itself. Therefore, it is important to explore the possibility of Majorana fermions in other spin-orbit coupled systems which are \emph{intrinsically} superconducting, obviating the need
  for the proximity effect. In this paper, we explore this possibility in non-centrosymmetric superconductors (NCS) where the superconductivity is inherent to the material. Gor'kov and Rashba have shown \cite{gorkov_rashba_2001} that strong spin-orbit interaction near a doped surface of a three-dimensional BCS-type superconductor lifts the two-fold spin degeneracy and the lack of inversion symmetry at the surface results in a two-dimensional pair wavefunction which is a mixture of both $s$-wave and $p$-wave parts. Penetration depth studies on NCS like $\tup{Li}_2\tup{Pd}_3\tup{B}$ and $\tup{Li}_2\tup{Pt}_3\tup{B}$ with purely $s$-wave interactions, have shown that the spin-orbit interaction and lack of inversion symmetry result in $s$-wave and $p$-wave mixing.~\cite{Yuan_Salamon_PRL_2006} The superconducting transition temperature ($T_c$) for Li$_2$Pd$_3$B is $6.7$ K and that for Li$_2$Pt$_3$B is $2.43$K.~\cite{Yuan_Salamon_PRL_2006} Similar admixture of $s$-wave and $p$-wave Cooper pair wavefunctions have been also been predicted in heavy fermion compounds like $\textup{CePt}_3\tup{Si}$ with broken inversion symmetry.~\cite{Frigeri_Sigrist_PRL_2004} A NCS with an externally applied Zeeman splitting is thus another candidate for realizing non-Abelian phases and Majorana bound states (MBS) in vortex cores.

However, due to the mixing of $s$-wave and $p$-wave pairing in NCS, it is {\it apriori} not obvious that the condition for the existence of Majorana modes 
as derived in Ref.~[\onlinecite{sau_prl_2010}] would hold even for NCS. Note that this condition, namely, $V_z^2 > \Delta_s^2 + \mu^2$ where $V_z, \Delta_s$ and $\mu$ are the Zeeman splitting, $s$-wave pair potential, and the chemical potential, respectively, is derived in Ref.~[\onlinecite{sau_prl_2010}] for the case when the $p$-wave pair potential is absent. In that case, since the superconducting pair potential is taken to be proximity-induced by an adjacent $s$-wave superconductor, \cite{sau_prl_2010} the question of a $p$-wave part of the pair potential does not arise.
 In this paper we will consider the spin-orbit coupled systems when both $s$-wave and $p$-wave pairing potentials are simultaneously present. While in some of the real systems the $s$-wave part of the pair potential has been found to be dominant over the $p$-wave component, in some other systems the $p$-wave part is dominant \cite{nishiyama}. By calculating the appropriate topological invariant and explicitly solving for the MBS using BdG equations, we shall show that exactly the same condition as stated above still holds for the existence of the non-Abelian state even in the presence of an explicit $p$-wave pair potential. This is important because it means that, irrespective of the relative values of the $s$-wave and the $p$-wave parts of the order parameter, a non-centrosymmetric superconductor can always be brought into a non-Abelian phase by the application of a sufficient Zeeman splitting. In contrast, existence of Majorana fermions in non-centrosymmetric superconductors was found earlier \cite{sato_fujimoto, lu_yip} only for the case when $|\Delta_t/\Delta_s|>1$, where $\Delta_t$ and $\Delta_s$ are the $p$-wave and the $s$-wave parts of the total pairing potential. 


The paper is organized as follows: In section I, we shall write down the BdG Hamiltonian for a 2D NCS with Zeeman splitting. We shall then give an explicit formula for the topological invariant and derive the condition for which the system is in a topologically non-trivial phase. In section II, we shall consider the non-degenerate Majorana bound states in vortex cores of such a NCS with Zeeman splitting. By writing down an effective 1D BdG equation and considering its solution out side the vortex core, we shall arrive at the condition for the MBS to exist. In section III, we shall consider the edge states in such a system and show that a non-degenerate MBS in vortex cores necessarily leads to gapless edge states. In section IV, we shall demonstrate the connection between the invariant in section I and the Pfaffian at zero momentum of the BdG Hamiltonian considered in section II. We shall summarize our findings in section V.

\section{Calculation of topological index for NCS with Zeeman splitting}
In this section we give an explicit formula for the topological invariant for non-centrosymmetric superconductor (NCS) in 2D and in the presence of a Zeeman splitting. The Hamiltonian for a two-dimensional NCS with both $s$-wave and $p$-wave pairing amplitudes and a Zeeman splitting applied in the $\hat{z}$ direction has the following form
\begin{widetext}
\begin{eqnarray}
H=\frac{1}{2}\sum_k (c_k^{\dagger}, \,\,c_{-k} ) \left(
\begin{array}{cc}
        \xi_k\sigma_0+V_z\sigma_z+\alpha(\sigbf\times \K).\hat{z}  & \Delta_s\sigma_0+\Delta_t(\sigbf\times \K).\hat{z} \\
        \Delta_s\sigma_0+\Delta_t(\sigbf\times \K).\hat{z} & \xi_k\sigma_0+V_z\sigma_z-\alpha(\sigbf\times \K).\hat{z}
\end{array}
\right) \left( \begin{array}{c}
c_k \\ c_{-k}^{\dagger}
\end{array}
\right)
\label{ham1}
\end{eqnarray}
\end{widetext}
 Here $\xi_k=\eta(k_x^2+k_y^2)-\mu$, $\eta=\hbar^2/2m$, $(c_k^{\dagger}, \,\,c_{-k} )=(c_{k,\uparrow}^{\dagger} \,\, c_{k,\downarrow}^{\dagger} \,\, c_{-k,\downarrow} \,\, -c_{-k,\uparrow})$, $V_z$ is a perpendicular Zeeman splitting which can be externally induced, $\alpha$ is the spin-orbit coupling, and $\Delta_s$ and $\Delta_t$ are the $s$-wave and the $p$-wave components of the pair potential, respectively. We note that the BdG Hamiltonian without time reversal and spin rotation symmetry belongs to class D \cite{Schnyder_classification_PRB_2008} and is characterized in 2D by a topological invariant $C_1$, which is the first Chern number of the $U(1)$ bundle describing the many-body wavefunction. The topological index $C_1$ written in terms of the quasiparticle Green's function $G(\K)$ has the following form:
\beqnar
C_1=\frac{1}{8\pi^2}\int d^2\K d\omega \left( \Tr[G\partial_{k_x}G^{-1}G\partial_{k_y}G^{-1}G\partial_{\omega}G^{-1}]\right.\nonumber \\ \left.-\Tr[G\partial_{k_y}G^{-1}G\partial_{k_x}G^{-1}G\partial_{\omega}G^{-1}]\right)
\label{Z-formula}
\eeqnar
Volovik and Yakovenko have derived the same expression for the topological invariant in $\tup{He}^3$-A \cite{volovik}.
 Writing the inverse Green's function as $G^{-1}=H-i\omega$, and using Eq.(\ref{ham1}) we numerically compute $C_1$ using the formula (\ref{Z-formula}) for a range of parameters. We find that the system is topologically non-trivial ($C_1=1$ with error bars $\pm 2\times 10^{-8}$) as long as the following condition is satisfied:
\begin{equation}
V_z^2>\mu^2+\Delta_s^2
\end{equation}
Note that this is the {\it same} constraint as obtained in Ref.~[\onlinecite{sau_prl_2010}] for the case when the $s$-wave pairing potential is proximity induced and, therefore, a triplet pair potential is absent. We thus find that the condition for the existence of non-Abelian phase remains unchanged even in the presence of triplet pairing and is thus completely independent of the ratio of triplet and singlet pairing amplitudes.

\begin{figure}
\begin{center}\vspace{.2cm}\end{center}
\includegraphics[width=3in]{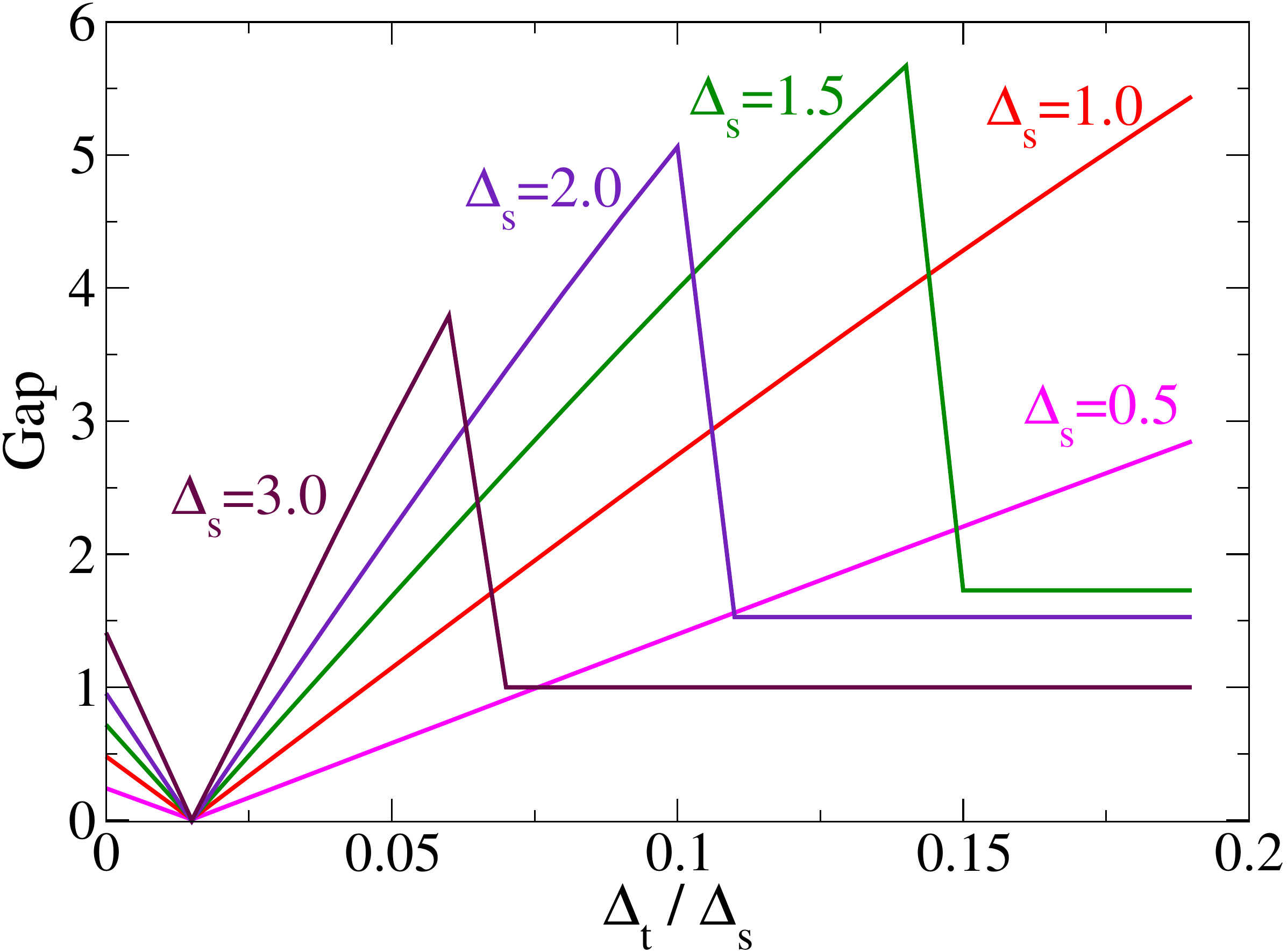}
\caption{Plot shows how the energy gap scales with the ratio $\Delta_t/\Delta_s$ for $V_z=6.0$, $\mu=4.0$, $\alpha=1.0$, and different values of $\Delta_s$ (all energy scales are in units of $\eta=\hbar^2/2m$). The discontinuity in the gap for $\Delta_s=3.0$, $\Delta_s=2.0$, and $\Delta_s=1.5$ are due to the global minimum of the gap function switching from $k\neq0$ to $k=0$ (see inset of Fig. \ref{gap}). Eventually, for very large values of $\Delta_s$, the global minimum always sits at $k=0$ and there is no gap closing at any value of $\Delta_t/\Delta_s$.}
\label{gap2}
\end{figure}

\begin{figure}
\begin{center}\vspace{.2cm}\end{center}
\includegraphics[width=3in]{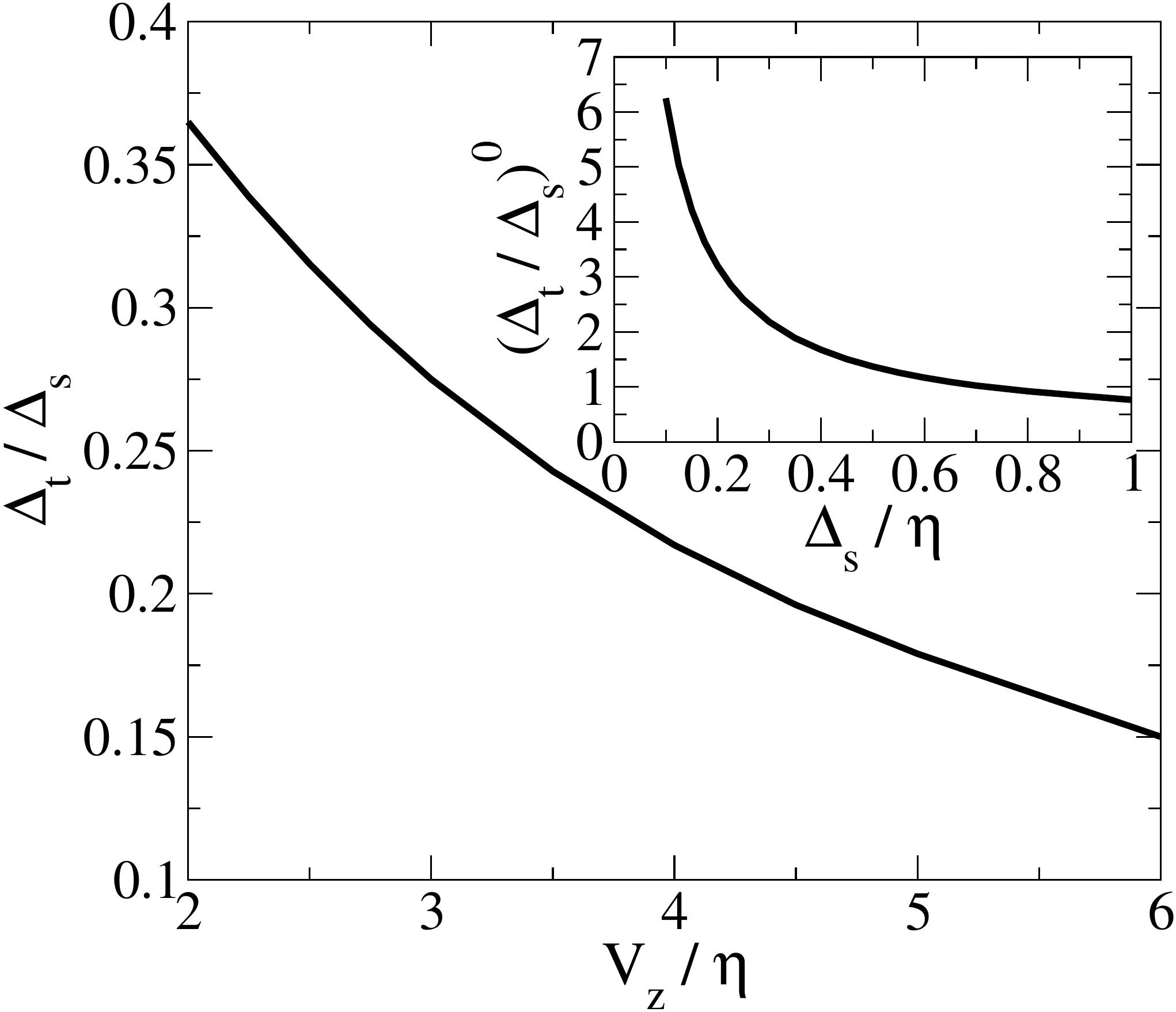}
\caption{Plot shows how the ratio $\Delta_t/\Delta_s$ at gap closing scales with the Zeeman splitting $V_z$ for $\alpha=1$, $\mu=1$, and $\Delta_s=0.5$ (in units of $\eta=\hbar^2/2m$). Note that, since $V_z^2>\mu^2+\Delta_s^2$ for the range of parameters in the plot, the topological invariant $C_1=1$ in spite of the gap closing. Inset shows the dependence on $\Delta_s$ of the ratio $(\Delta_t/\Delta_s)_0$ at which global minimum of the gap shifts from $k_x, k_y\neq 0$ to the origin. For large $\Delta_s$, the global minimum is always at $k_x=k_y=0$. Following values have been used: $V_z=6.0$, $\mu=4.0$, and $\alpha=1.0$ (in units of $\eta=\hbar^2/2m$)}
\label{gap}
\end{figure}

This result is a major improvement over Ref.~[\onlinecite{sato_fujimoto}] where only the case $|\Delta_t/\Delta_s|>1$ has been considered. In that work it has been claimed that, in order for the non-Abelian phase to be stable, the single particle energy gap should not close. A sufficient condition for the single particle gap to not close is $|\Delta_t/\Delta_s|>1$. This condition for the existence of the non-Abelian phase is required if the Chern number is calculated (as is done in Ref.[\onlinecite{sato_fujimoto}]) by decoupling the spins in the limit $\alpha \to 0, \Delta_s \to 0$ and mapping the system to two independent copies of spinless $p_x+ip_y$ superconductors. One can then calculate the TKNN numbers \cite{TKNN} for the two spins independently provided the gap does not close during the mapping, and this gives rise to the unnecessary additional constraint $|\Delta_t/\Delta_s|>1$ for the existence of the non-Abelian phase. Since our method of calculating the index does not rely on this mapping, we find quite generally that the non-Abelian phase survives even if the gap closes at some value of $|\Delta_t/\Delta_s|$ and momentum $|\K|$. This is illustrated in Fig. \ref{gap2} and Fig. \ref{gap}.

As is clear from Fig.~\ref{gap2}, the excitation gap, which is defined as the momentum-space global minimum of the lowest energy positive eigenvalue 
of the Hamiltonian in Eq.~(\ref{ham1}), starts being non-zero for small values of the ratio $\Delta_t/\Delta_s$ (we have chosen parameters such that
the condition $V_z^2 > \Delta_s^2 + \mu^2$ is satisfied). The gap decreases with increasing values of $\Delta_t/\Delta_s$, eventually vanishing at some (parameter-dependent) value of this ratio. However, for larger value of $\Delta_t/\Delta_s$ the gap opens again.  One of the principal finding of this paper is that, even though the gap closes at some value of $\Delta_t/\Delta_s$ (and $|\K|\neq 0$), the index $C_1=1$ on both sides of the gap closing point. This indicates that the non-Abelian phase survives on both sides of the gap closing point on the $\Delta_t/\Delta_s$ axis. Therefore, the closing of the gap with increasing values of $\Delta_t$ is \emph{not} associated with a topological quantum phase transition.
In Fig.~\ref{gap2}, for $\Delta_t/\Delta_s$ larger than the value for which the gap closes, the gap also shows a discontinuity. This is because it discontinuously shifts from being at a nonzero $k$ to $k=0$. We find that, for a given set of values for $\alpha, V_z, \mu,$ and $\Delta_s$, the gap lies at $|\K|\neq 0$ for small values of the ratio $\Delta_t/\Delta_s$. Beyond a certain value of $\Delta_t/\Delta_s$ (which is parameter-dependent  and is typically larger than the value for which the gap closes), the gap shifts to $|\K|= 0$ and, as illustrated in Fig.~\ref{gap2}, is discontinuous when this shift occurs. In Fig.~\ref{gap}, we show the various parameter dependencies of the two special values of $\Delta_t/\Delta_s$, the one at which the gap closes, and the one at which the gap shifts in the momentum space and is discontinuous.  

It is worth reiterating the surprising result stated above (which we confirm below also by explicit real space solutions of the BdG equations): even though the excitation gap of the superconducting system does vanish at a certain value of $\Delta_t/\Delta_s$ and is non-zero on either side of this axis, the system has precisely the same non-Abelian topological properties on both sides of the gap closing point. This is a consequence of the fact that gap closing is a necessary but not a sufficient condition for having a topological phase transition. The superconducting states on both sides of the gap closing point are characterized by order parameter defects carrying zero-energy non-Abelian Majorana fermion excitations. 
The only topological phase transition that takes place in this system happens when the chemical potential or the Zeeman splitting are tuned to values such that $V_z^2=\mu^2+\Delta_s^2$.

The integer computed in Eq.~(\ref{Z-formula}) is essentially the same as the TKNN number which gives rise to quantized Hall conductivity and topologically protected edge currents in quantum Hall systems.~\cite{TKNN} There is a one-to-one correspondence between quantum Hall systems and  superconductors with broken time reversal symmetry - the BdG Hamiltonian characterizing the quasiparticles in the latter also describes quantum Hall insulators, with the superconducting gap being replaced by the bulk insulating gap. We shall explicitly show the connection between $C_1$ and the TKNN invariant in section IV.

\section{Existence of Majorana modes in NCS with Zeeman splitting}
In this section we shall explicitly derive an existence condition for the non-degenerate zero-energy Majorana modes in the vortex core of a NCS with Zeeman splitting. In the absence of a Zeeman splitting, time reversal invariance guarantees two degenerate Majorana modes bound to vortex cores. 
We shall show that the existence criterion for single Majorana fermions bound to vortex cores is the same condition derived in the previous section. The BdG Hamiltonian in the presence of a vortex with winding number $n$ can be written in real space using polar co-ordinates as
\begin{widetext}
\beqnar
H_{\textup{BdG}}=[H_0\tau_z\,+\,\Delta_s(r)\tau_++h.c.]\delta(r-r')+\frac{1}{2}\left[ \Delta_t(r,r')\sigma_+ + \Delta_t^*(r,r')\sigma_-\right]\left[ \cos(n\theta)\tau_x+\sin(n\theta)\tau_y\right]
\label{bdg1}
\eeqnar
\end{widetext}
In the above equation $H_0=-\hbar^2\nabla^2/2m-\mu+V_z\sigma_z+\alpha({\bf \sigma}\times \Pmom).\hat{z}$. For compactness we shall write the Hamiltonian in the momentum space where the $p$-wave gap is given by $-i\Delta_t({\bf \sigma}\times \Pmom).\hat{z}[\cos(n\theta)\tau_x+\sin(n\theta)\tau_y]$. The Hamiltonian (\ref{bdg1}) can be diagonalized by noting that one can construct a pseudo-angular momentum operator $J_z=L_z+(\sigma_z-n\tau_z)/2$ that commutes with the Hamiltonian. One can therefore use a canonical transformation ${\tilde H}_{\textup{BdG}}=e^{-ij_z\theta}H_{\textup{BdG}}e^{ij_z\theta}$ to make the Hamiltonian independent of $\theta$, where $j_z=m_j-\sigma_z/2+n\tau_z/2$ and $m_J$ are the eigenvalues of the operator $J_z$. We next note that under a particle-hole transformation $m_J\to-m_J$. Since we are interested in the Majorana solution we therefore set $m_J=0$. To further simplify the analysis we focus on vortices with unit winding number $n=1$ and make another canonical transformation $e^{-i\sigma_z\pi/4}{\tilde H}_{\textup{BdG}}e^{i\sigma_z\pi/4}$ that makes the Hamiltonian real. The $E=0$ solution of the resulting BdG Hamiltonian would in general yield two solutions that are complex conjugate pairs: $\Psi(r)$ and $\Psi^*(r)$. Also, due to the particle-hole symmetry of the BdG Hamiltonian, if $\Psi(r)$ is a solution then so is $\sigma_y\tau_y\Psi(r)$. If these two solutions are related by $\Psi(r)=\lambda\sigma_y\tau_y\Psi(r)$, then $(\sigma_y\tau_y)^2=1$ implying $\lambda=\pm 1$. As a final step, we use the relation $\sigma_y\tau_y=\lambda$ to decouple the particle and hole sectors in the pairing terms by writing $\tau_x=i\lambda\sigma_y\tau_z$. The pairing terms are then given by $(i\lambda \sigma_y \tau_z)[\Delta_s-i\Delta_t({\bf \sigma}\times \Pmom).\hat{z}]$ and the particle-hole sectors given by $\tau_z=\pm1$ are thus completely decoupled.

The BdG equations for the $n=1$ vortex with zero energy written in real space is then given by
\begin{widetext}
\beqnar
\left(
\begin{array}{cc}
        -\eta\left( \partial_r^2+{1\over r}\partial_r\right)+V_z-\mu-\Delta_t\partial_r  & \lambda\Delta_s(r)+\alpha\left(\partial_r+{1\over r}\partial_r \right) \\
        -\lambda\Delta_s(r)-\alpha\partial_r  & -\eta\left( \partial_r^2+{1\over r}\partial_r-{1\over r^2}\right)-V_z-\mu-\Delta_t\left(\partial_r-{1\over r}\partial_r \right)
\end{array}
\right)\Psi_0(r)=0
\eeqnar
\end{widetext}
where $\eta=\hbar^2/(2m)$.

\begin{figure}[h]
\begin{center}$
\begin{array}{ccc}
\includegraphics[width=1.4in]{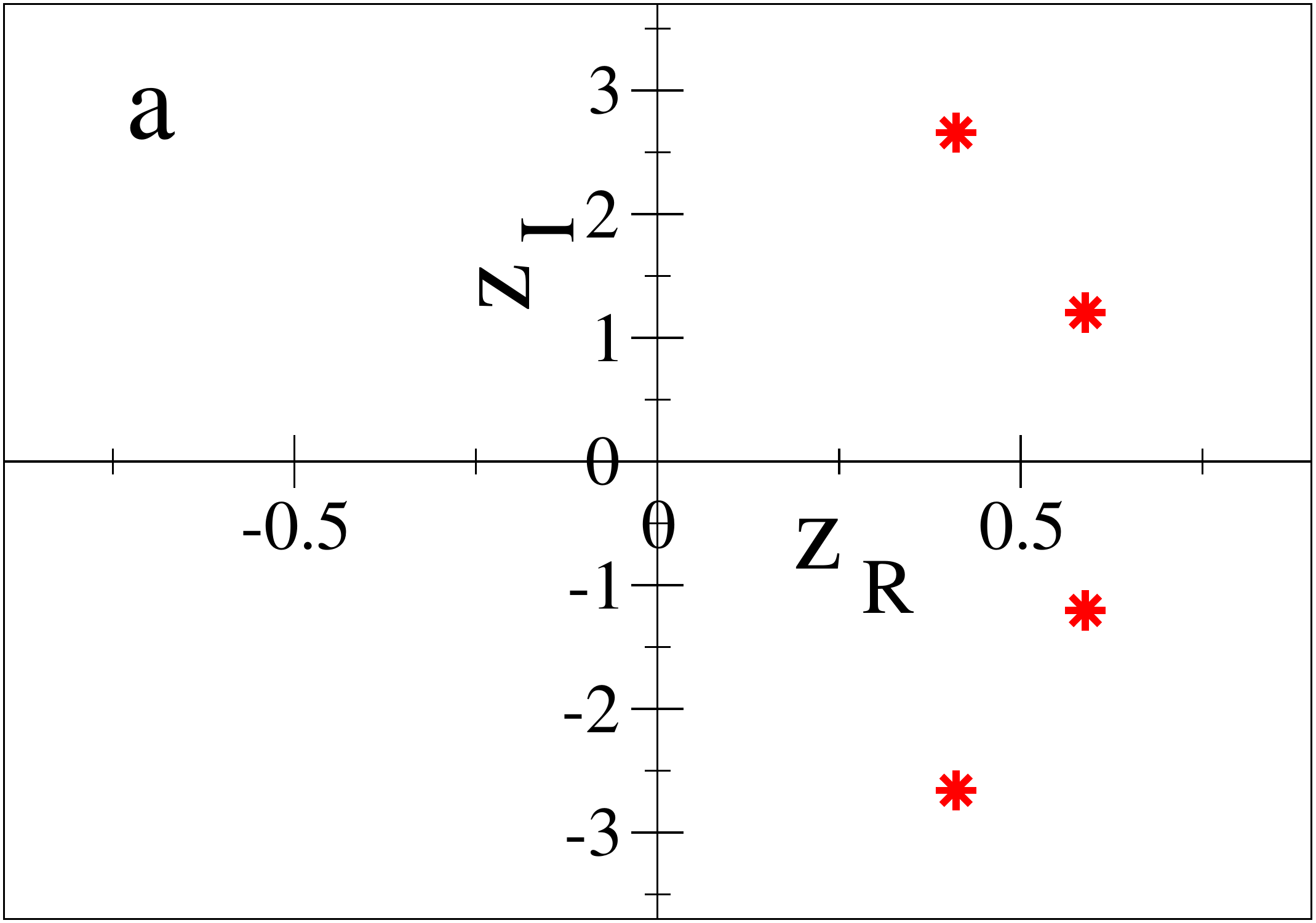}&\,\,\,\,&
\includegraphics[width=1.4in]{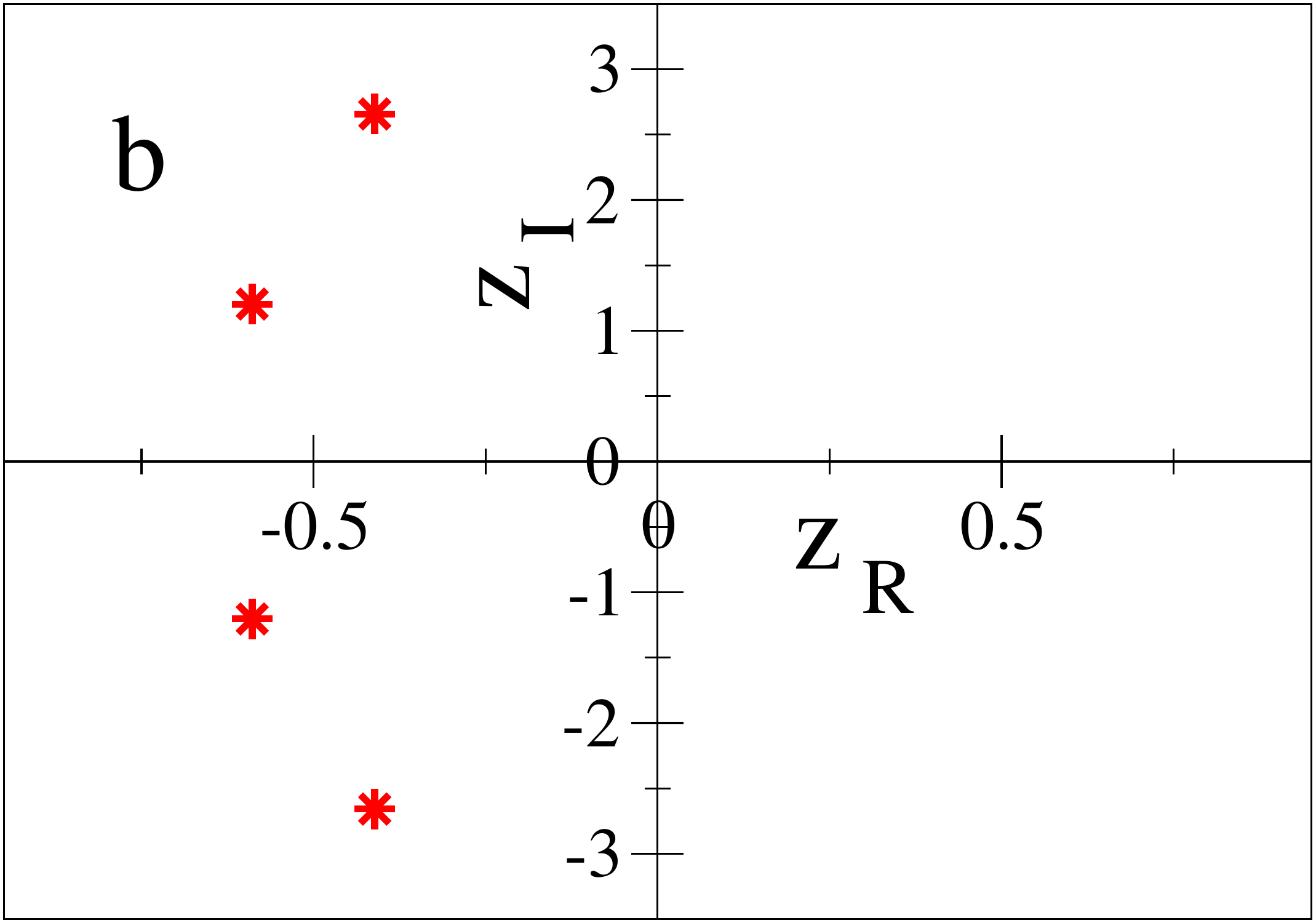} \\\\\\
\includegraphics[width=1.4in]{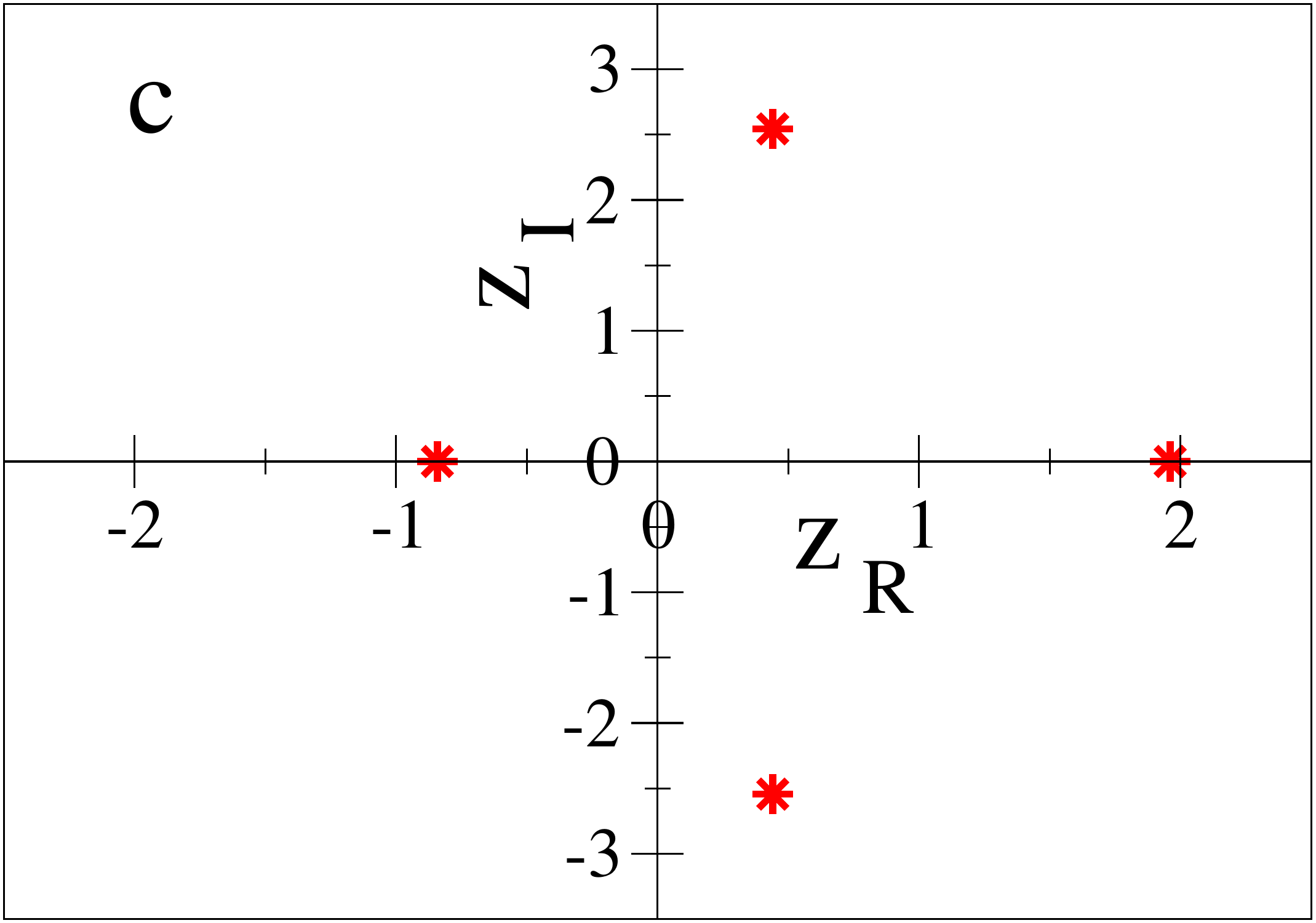}&\,\,\,\,&
\includegraphics[width=1.4in]{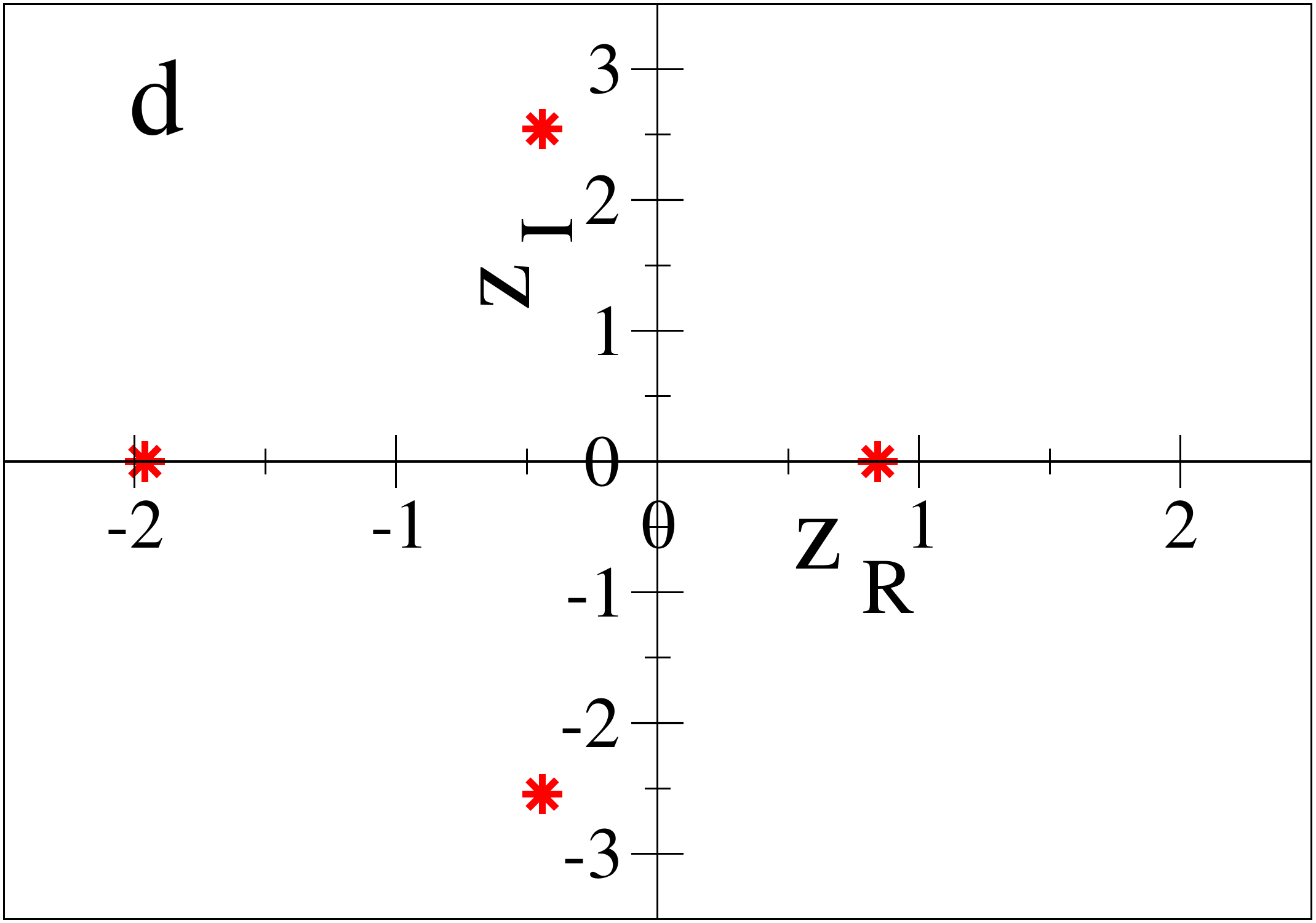}
\end{array}$
\end{center}
\caption{Nature of roots for Eq. (\ref{vortex_eqn}) for the following parameters: $\alpha=1$, $\Delta_s=1$, $\Delta_t=1$, and (a) $\mu=4$, $V_z=2$, $\lambda=+1$, (b) $\mu=4$, $V_z=2$, $\lambda=-1$, (c) $\mu=2$, $V_z=4$, $\lambda=+1$, (d) $\mu=2$, $V_z=4$, $\lambda=-1$ (in units of $\eta=\hbar^2/2m$)}
\label{roots}
\end{figure}

Following \cite{sau_prl_2010} we next approximate the radial dependences of $\Delta_s(r)$ and $\Delta_t(r)$ by $\Delta_{s,t}(r)=0$ for $r<R$, where $R$ is the size of the vortex core and $\Delta_{s,t}(r)=\Delta_{s,t}^0$ for $r\geq R$. Since both the order parameters are zero inside the vortex core, one can construct explicit analytical solutions in terms of Bessel functions. The characteristic equation leads to two linearly independent solutions that correspond to the two Fermi surfaces obtained by the intersection of the two bands with the Fermi level. Out side the vortex, we seek solutions of the type:
\beqnar
\left(
\begin{array}{c}
u_{\upar}(r) \\ u_{\dnar}(r) \end{array}\right)={e^{zr}\over r^{1/2}}\left( \begin{array}{c}\rho_{\upar}(1/r) \\ \rho_{\dnar}(1/r) \end{array}\right)
\label{vortex_soln}
\eeqnar
such that $\rho_{\sigma}(x)$ are analytic functions of $x$. This leads to a convergent power-series solution in $1/r$, whose zeroth order term can be obtained by setting $1/r=0$ which leads to the following equation:
\beqnar
\left(
\begin{array}{cc}
        \xi_z+V_z+z\Delta_t  & \lambda\Delta_s-z\alpha \\
        -\lambda\Delta_s+z\alpha  & \xi_z-V_z+z\Delta_t
\end{array}
\right)\left( \begin{array}{c}\rho_{\upar}(0) \\ \rho_{\dnar}(0) \end{array}\right)=0
\eeqnar
where $\xi_z=-\eta z^2-\mu$. The two families of solutions for $\lambda=\pm 1$ are related to each other by $z\to-z$. The values of $z$ that are consistent with the above equation are obtained by solving
\beq
(\xi_z\pm z\Delta_t)^2-V_z^2+(z\alpha\mp \Delta_s)^2=0
\label{vortex_eqn}
\eeq
The sign of the product of the roots of this quartic equation is given by $S=\sgn(\Pi_n(z_n))=\sgn(\Delta_s^2+\mu^2-V_z^2)$. When $\Delta_s^2+\mu^2-V_z^2>0$ there are 2 real roots and 2 complex roots, whereas when $\Delta_s^2+\mu^2-V_z^2<0$ there are 4 complex roots. The different possibilities are enlisted in Table-1.
\begin{table}[ht]
\centering
\begin{tabular}{|c|c|c|c|}
\hline\rule[-.4cm]{0cm}{1cm}
{\bf sgn($\mathbf{C_0}$)}&$\,\,\,{\boldsymbol \lambda}\,\,\,$&{\bf \em{Nature of Roots}}&${\bf Re(z_n)<0}$
\\\hline\rule[-.4cm]{0cm}{1cm}
$+$&$\,\,\,+1\,\,\,$&4 Complex roots& For no root\\\hline\rule[-.4cm]{0cm}{1cm}
$+$&$\,\,\,-1\,\,\,$&4 Complex roots& For all 4 roots\\\hline\rule[-.4cm]{0cm}{1cm}
$-$&$\,\,\,+1\,\,\,$&2 Real, 2 Complex roots& For 1 root\\\hline\rule[-.4cm]{0cm}{1cm}
$-$&$\,\,\,-1\,\,\,$&2 Real, 2 Complex roots& For 3 roots\\\hline
\end{tabular}
\caption{Summary of solutions for Eq. (\ref{vortex_eqn}) for $\sgn(C_0)=\pm$ and $\lambda=\pm 1$.}
\label{phi-notation}
\end{table}\newline
Figure (\ref{roots}) show the roots of Eq. (\ref{vortex_eqn}) for $\sgn(C_0)=\pm$ and $\lambda=\pm 1$ and a given set of parameters.
Below we show that in the first three cases, there is no unique solution for $\Psi(r)$, whereas in the fourth case an unique zero-energy Majorana solution exists.

In order for the solution (\ref{vortex_soln}) to be normalizable we need $Re(z_n)<0$ for $r>R$. Further, an unique solution for the wavefunction $\Psi(r)$ of the zero-energy non-degenerate Majorana mode, would require the two-component wavefunctions $\Psi(r<R)$ and $\Psi(r>R)$ to satisfy four boundary conditions for $\Psi(r)$ and $\Psi'(r)$ at the boundary $r=R$. One additional condition comes from the normalization condition for $\Psi(r)$. This leads to a total of five equations for $\Psi(r)$ and hence an unique solution (\ref{vortex_soln}) exists when there are exactly five unknown co-efficients in $\Psi(r)=\sum_n c_n \Psi_n(r)$ corresponding to the roots $z_n$ of Eq. (\ref{vortex_eqn}). From the discussion in the previous paragraph, we find that when $\Delta_s^2+\mu^2-V_z^2<0$ and $\lambda=+1$, then $\Psi(r<R)=\sum_{n=1,2} c_n \Psi_n(r)$ and $\Psi(r>R)=\sum_{n=1,2,3} c_n \Psi_n(r)$ and hence there are exactly five unknown co-efficients. Hence, there is an unique zero-energy Majorana solution in this case. On the other hand, when $\Delta_s^2+\mu^2-V_z^2>0$, one can easily check that either there are no normalizable solutions or the number of unknown co-efficients do not equal to the number of equations and hence no unique zero-energy solution exists.

\section{Non-Chiral edge modes in NCS with Zeeman splitting}
It is well known that topological phases are characterized by the existence of gapless edge states. Here we shall explicitly demonstrate the existence of edge states in NCS with Zeeman splitting and how the edge states are related to the existence of Majorana modes in vortex cores discussed in the previous section. Without loss of generality we consider an edge perpendicular to $\hat{y}$. The Hamiltonian, which is now a function of $k_x$ can be shown to be symmetric under $k_x\to -k_x$. This property of the Hamiltonian follows from particle-hole symmetry. Therefore, in analogy with the calculation of the Majorana modes we set $k_x=0$. The resulting Hamiltonian is then a single band 1D BdG Hamiltonian with spin-orbit coupling given by
\beqnar
H_{\textup{BdG}}=&\hspace{-1.6cm}[-\eta \partial_y^2-\mu(y)+V_z \sigma_z-i\alpha\sigma_x\partial_y]\tau_z\nonumber \\&\hspace{-.2cm}+(\Delta_s(y)\tau_+ + h.c.)+\Delta_t(y)(-i\sigma_x\partial_y)(i\sigma_y\tau_z)
\eeqnar
We note that the non-degenerate spinor solution has the Majorana form $\Psi=(u, i\sigma_y u^*)$, where $u$ is a two spinor. Writing $u=u_R+iu_I$, one obtains a pair of equations in which $u_R$ and $u_I$ are decoupled and are completely analogous to the $\lambda=\pm 1$ channels for the Majorana case. One can define the edge at $y=0$ as $\mu(y<0)=\mu$ and $\mu(y>0)=0$, and look for solutions of the type $\Psi(y)=\sum_n a_n e^{-z_n y}$. The zero energy mode of the Hamiltonian written in the reduced $2\times 2$ space (say $\lambda=+1$ channel) then takes the form
\beqnar
\left(
\begin{array}{cc}
        \xi_{z_n}+V_z+z\Delta_t  & \lambda\Delta_s-z_n\alpha \\
        -\lambda\Delta_s+z_n\alpha  & \xi_{z_n}-V_z+z_n\Delta_t
\end{array}
\right)u_n=0
\eeqnar
where $\xi_{z_n}=-\eta z_n^2-\mu$. This is the same equation as obtained for the solution outside the vortex core. Matching the boundary conditions and normalizing the solution yields five equations which bear an unique solution when the number of unknowns is also five. Following the analysis in the previous section one can see that the latter condition is satisfied when $V_z^2>\mu^2+\Delta_s^2$.

\section{Pfaffian $Z_2$ invariant and Chern number invariant for BdG Hamiltonians.}
In the previous sections, we characterized the time-reversal symmetry breaking
non-centrosymmetric superconductors first by the first Chern number
topological invariant $C_1$ and then by the sign
 $P=\sgn{(\mu^2+\Delta_s^2-V_Z^2)}$.
The first Chern number $C_1$ is an integer $(\mathbb{Z})$ invariant
 while the latter quantity is a $Z_2$ invariant that
 characterizes whether the system is in a phase that supports Majorana
fermions or not at defects.
 Numerical calculation of $C_1$ suggests that both
$P=\pm 1$ and $C_1=0,1$ are consistent characterization of when the
system supports a Majorana Fermion. In this section, we explicitly
show that $P$ is a special case of a Pfaffian invariant for
 $H_{BdG}(\bm k=0)$, which is shown to be equal to the parity of
the first Chern number $C_1$, analogous to the case of topological
insulators \cite{fukane_PRB}.

\subsection{Pfaffian Invariant}
The $\bm k=(k_x,k_y)$ dependent Bogoliubov-de Gennes Hamiltonian $H_{BdG}(\bm k)$ has a well-known particle-hole symmetry which is written as
\begin{equation}
\Xi H_{BdG}(\bm k)\Xi^{-1}=\Lambda H_{BdG}^*(\bm k)\Lambda=-H_{BdG}(-\bm k)\end{equation}
where $\Xi=\Lambda K$ and $\Lambda=(\sigma_y \tau_y)$ and $K$ is the
complex conjugation operator. Here $\sigma_{x,y,z}$ are the Pauli spin matrices and $\tau_{x,y,z}$ are the Nambu particle-hole matrices.

The particle-hole symmetry $\Xi$, similar to the time-reversal symmetry,
maps $\bm k\rightarrow -\bm k$. Therefore similar to time-reversal
invariant topological insulators, some of the topological properties
of particle-hole symmetric BdG Hamiltonians can be extracted from
the Hamiltonian at particle-hole symmetric k-points such that
$\bm k=-\bm k+\bm G$ where $\bm G$ is a reciprocal lattice vector. We
refer to such points as $\bm K$. The BdG Hamiltonian at such points
$H_{BdG}(\bm K)$ is explicitly particle-hole symmetric.
Given such a Hamiltonian, we can define a matrix $W(\bm K)$
which can be shown to be  anti-symmetric by virtue of the particle-hole symmetry as follows:
\begin{equation}
W^T(\bm K)=\Lambda^T H^T(\bm K)=-\Lambda\Lambda H(\bm K)\Lambda=-W(\bm K).
\end{equation}

Moreover one can show that $i^n Pf(W)$ is real since
\begin{align*}
&Pf(W^*(\bm K))=Pf(W(\bm K))^*=Pf(H_{BdG}^*(\bm K)\Lambda)\\
&=Pf(H^T_{BdG}(\bm K)\Lambda)=\frac{Pf(\Lambda^T H^T_{BdG}(\bm K) \Lambda^2)}{Det(\Lambda)}\\
&=Pf(W^T(\bm K))=(-1)^n Pf(W(\bm K))
\end{align*}
where $H_{BdG}(\bm K)$ is a $2 n\times 2 n$ Hermitian matrix.
Using this one can define the sign
\begin{equation}
Q(H_{BdG}(\bm K))=\sgn{(i^n Pf(H_{BdG}(\bm K)\Lambda))}
\end{equation}
for any particle-hole symmetric BdG Hamiltonian.
The function $Q$ defined above can change only if
 $Pf(H_{BdG}(\bm K)\Lambda)$ vanishes.
Since $$Pf(H_{BdG}(\bm K)\Lambda)^2=Det(H_{BdG}(\bm K)\Lambda)=Det(H_{BdG}),$$ this implies that $Q$ can only change sign where the gap of the BdG
 Hamiltonian vanishes at $\bm K$. Therefore $Q$ defines a topological invariant for the space of particle-hole symmetric BdG Hamiltonians.
Note also that this topological invariant does not rely on any other symmetry such as time-reversal symmetry. The invariant $Q$ is related to the invariant suggested by Kitaev for one-dimensional $p_x$ superconductors \cite{Kitaev}. It can be easily verified that for the Hamiltonian in Eq.(\ref{ham1}), the invariant $Q(H(\bm 0))=\sgn{(V_Z^2-\Delta_s^2-\mu^2)}$.

The Pfaffian invariant $Q$ can be computed in terms of the eigenvectors of $H_{BdG}(\bm K)$.
 Suppose $H_{BdG}(\bm K)$ is diagonalized by the transformation
$$H_{BdG}(\bm K)=U(\bm K)D(\bm K)U^\dagger(\bm K)$$
where $D(\bm K)$ is a diagonal matrix of eigenvalues ordered in descending order of value.
The columns of the unitary matrix $U(\bm K)$ are the eigenvectors of  $H_{BdG}(\bm K)$. The
positive energy eigenvectors in $U$ are chosen to be related to the negative energy eigenvectors
by particle-hole symmetry. With such a convention for the eigenvectors and eigenvalues,
the unitary matrix $U(\bm K)$ satisfies the constraint
$$\Lambda U(\bm K)=U^*(\bm K)\Gamma$$
 where $$\Gamma=\left(\begin{array}{cc}0&I_n\\I_n&0\end{array}\right)$$
and $I_n$ is the $n\times n$ identity matrix. The Pfaffian of $H_{BdG}(\bm K)\Lambda$ is then
\begin{equation}
Pf(H_{BdG}(\bm K)\Lambda)=Det(U(\bm K))Pf(D(\bm K)\Gamma)
\end{equation}
where $Pf(D(\bm K)\Gamma)=\prod_{n>0}E_n(\bm K)$. Therefore the topological invariant $Q$
is determined by the sign of $Det(U(\bm K))$ as
\begin{equation}
Q(H_{BdG}(\bm K))=Det(U(\bm K)).
\end{equation}
The above formula can be used to evaluate $Q(H_{BdG}(\bm K))$ at the particle-hole invariant points
$\bm K=(0,\pm\frac{\pi}{a}),(\pm\frac{\pi}{a},0),(\pm\frac{\pi}{a},\pm\frac{\pi}{a})$ using the fact
that $\epsilon_{\bm K}\gg V_Z,\Delta_s,\mu$. To dominant order in $a$, $H_{BdG}(\bm K)=\epsilon_{\bm K}\tau_z$.
Therefore at these points $Q(H_{BdG}(\bm K))=1$.

\subsection{Relation to Berry connection.}
The evolution of the eigenvector matrix $U(k_x,k_y)$
with $k_x$ can be computed
by introducing the multi-band vector potential  $$\bm A_{m,n}(k_x,k_y)=-i\langle u_{m}(k_x,k_y)|\bm\nabla |u_{n}(k_x,k_y)\rangle.$$
 The evolution of the eigenvector matrix can be
written in terms of this potential as
\begin{align*}
&\partial_{k_x} U(k_x,k_y)=i U(k_x)A^{(x)}(k_x,k_y)\\
&U(k_x,k_y)^{-1}\partial_{k_x} U(k_x,k_y)=i A^{(x)}(k_x,k_y)\\
&\partial_{k_x} \log(U(k_x,k_y))=i A^{(x)}(k_x,k_y)\\
&\partial_{k_x} \log(Det(U(k_x,k_y)))=i Tr(A^{(x)}(k_x,k_y))\\
&P(k_y=0,\pi)=\frac{Det(U(k_x=\pi,k_y))}{Det(U(k_x=0,k_y))}=e^{i \pi S}.
\end{align*}
Here $S$ is integral of the trace and is calculated as
\begin{align*}
&S=\frac{1}{\pi}\int_0^\pi d k_x Tr(A^{(x)}(k_x,k_y))\\
&=\int_0^\pi \frac{d{k_x}}{\pi} \sum_{m}\langle u_{m}(k_x,k_y)|\partial_{k_x} |u_{m}(k_x,k_y)\rangle.
\end{align*}
This integral by itself does not reduce to a BZ integral since the
integral over $k_x$ extends only to $\pi$. Using particle hole symmetry
of the BdG Hamiltonian it is possible to
pick Bloch states at $-k_x$ such that
\begin{equation}
u_{-m}(-k_x,k_y=0,\pi)=\sigma_y\tau_y u^*_{m}(k_x,k_y=0,\pi)
\end{equation}
where $-m$ is the state in the band $-m$.
In this gauge, the connection away from $k_x=\pi$ must satisfy
\begin{align*}
&u_m(k_x,k_y)^\dagger\partial_{k_x} u_m(k_x,k_y)=-u_m(k_x,k_y)^T \partial_{k_x} u_{m}(k_x,k_y)\\
&=-u^\dagger_{-m}(-k_x,k_y)\partial_{k_x} u_{-m}(-k_x,k_y)\\
&=u^\dagger_{-m}(k'_x,k_y)\partial_{k'_x}u_{-m}(k_x,k_y)|_{k'_x=-k_x}.
\end{align*}

Therefore the integral of the trace $S$ becomes
\begin{align*}
&S=\int_0^\pi d{k_x} \sum_{m<0}\langle u_m(k_x,k_y)|\partial_{k_x} |u_m(k_x,k_y)\rangle\\
&+\int_0^\pi d{k_x} \sum_{m>0}\langle u_m(k_x,k_y)|\partial_{k_x} |u_m(k_x,k_y)\rangle\\
&=\int_0^\pi d{k_x} \sum_{m<0}\langle u_m(k_x,k_y)|\partial_{k_x} |u_m(k_x,k_y)\rangle\\
&+\int_{-\pi}^0 d{k_x} \sum_{m<0}\langle u_m(k_x,k_y)|\partial_{k_x} |u_m(k_x,k_y)\rangle\\
&=\int_{-\pi}^\pi dk\sum_{m<0}\langle u_m(k_x,k_y)|\partial_{k_x} |u_m(k_x,k_y)\rangle.
\end{align*}
The integral $S$ can be written in terms of a $U(1)$ vector potential $\bm a(k_x,k_y)$
\begin{equation}
S(k_y)=\int_{-\pi}^{\pi} \frac{d k_x}{\pi} a_x(k_x,k_y)
\end{equation}
where
\begin{equation}
\bm a(k_x,k_y)=\sum_{m<0} \langle u_m(k_x,k_y)| i \nabla | u_m(k_x,k_y) \rangle.
\end{equation}

The line integral over $k_x$ can be converted to a loop integral that encloses a finite area of the
BZ torus by subtracting the integrals at $k_y=0$ and $k_y=\pi$ i.e.
\begin{align}
&S=S(k_y=\pi)-S(k_y=0)\nonumber\\
&=\int_{-\pi}^{\pi} d k_x a_x(k_x,k_y=\pi)-\int_{-\pi}^{\pi} d k_x a_x(k_x,k_y=0)\nonumber\\
&=\oint d k_x a_x(k_x,k_y).
\end{align}
Using Stokes theorem this integral can be written in terms of the
Berry curvature $f_{xy}=\partial_{k_x}a_{k_y}-\partial_{k_y}a_{k_x}$ as
\begin{equation}
S=\oint d k_x a_x(k_x,k_y)=\int dk_x d k_y f_{xy}(k_x,k_y)
\end{equation}
which as pointed out by TKNN \cite{TKNN,qihugheszhang}
is written as
\begin{align}
&S=\int_{-\pi}^{\pi} dk_x\int_{0}^{\pi} d k_y \sum_{m,n}(f_m-f_n)\nonumber\\
&\frac{\langle u_{m}|\partial_{k_x}H_{BdG}|u_{n}\rangle\langle u_{n}|\partial_{k_y}H_{BdG}|u_{m}\rangle}{(E_m-E_n)^2}
\end{align}
where $f_{m,n}$ are the Fermi occupation functions.

Using particle-hole symmetry, the above equation can be extended to the entire BZ to yield the final result
\beqnar
P=\frac{Q(H_{BdG}(k_x=0,k_y=0))Q(H_{BdG}(k_x=\pi,k_y=\pi))}{Q(H_{BdG}(k_x=\pi,k_y=0))Q(H_{BdG}(k_x=0,k_y=\pi))}\nonumber \\ =e^{i \pi S}\hspace{7cm}
\eeqnar
where $S=\int_{BZ} d^2 k f_{xy}$
and
\begin{equation}
f_{xy}=\sum_{m,n}(f_m-f_n)\frac{\langle u_{m}|\partial_{k_x}H_{BdG}|u_{n}\rangle\langle u_{n}|\partial_{k_y}H_{BdG}|u_{m}\rangle}{(E_m-E_n)^2}.
\end{equation}

Since, as argued in the previous sub-section, the Q invariant at the
particle-hole symmetric points other than $\bm K=0$ are 1, the results of this
sub-section shows that the Pfaffian topological invariant $Q(H_{BdG}(k_x=0,k_y=0))$ is related to the TKNN number $S$ by
\begin{equation}
Q(H_{BdG}(k_x=0,k_y=0))=e^{\imath \pi S}.
\end{equation}
This is the central result of this section.

\subsection{Connection to Green function form of invariant.}
The TKNN invariant can be expressed compactly in terms of the Green
function as\cite{volovik}
\begin{align}
&C_1=\int d^2 k d\omega Tr[G\partial_{k_x}G^{-1}G\partial_{k_y}G^{-1} G\partial_{\omega}G^{-1}]\nonumber\\
&- Tr[G\partial_{k_y}G^{-1}G\partial_{k_x}G^{-1} G\partial_{\omega}G^{-1}]
\end{align}
where $G^{-1}(k_x,k_y,\omega)=\omega-H(k_x,k_y)$. To see that this is indeed the
same as the TKNN expression, we note first that the above equation is
simply
\begin{align}
&C_1=\int d^2 k d\omega Tr[G^2\partial_{k_x}G^{-1}G\partial_{k_y}G^{-1}]\nonumber\\
&- Tr[G^2\partial_{k_y}G^{-1}G\partial_{k_x}G^{-1}].
\end{align}
Expanding the Green function in terms of eigenstates we see that the
frequency integral in the momentum can be written as
\begin{align*}
&\int d\omega \sum_{m,n}\frac{1}{(\omega-E_m)^2}\langle u_{m}\partial_{k_x}H|u_n\rangle \frac{1}{\omega-E_n}\langle u_n|\partial_{k_y}H|u_m\rangle\\
&=\sum_{m,n}\frac{(f_m-f_n)}{(E_m-E_n)^2}\langle u_{m}\partial_{k_x}H|u_n\rangle \langle u_n|\partial_{k_y}H|u_m\rangle=f_{xy}.
\end{align*}

The invariant $C_1$ then becomes the TKNN invariant
\begin{equation}
C_1=S=\int_{BZ} d^2 k f_{xy}(k_x,k_y).
\end{equation}
\section{Conclusion}
We have considered non-centrosymmetric superconductors in the presence of a Zeeman splitting normal to the surface. The resulting two-dimensional superconductor on the surface has both $s$-wave and $p$-wave pairing. We have derived the condition for which the system is in a non-Abelian phase by constructing a topological invariant. We find that the condition for a non-Abelian phase to exist is completely independent of the triplet pairing potential. This is in contrast to the case of time reversal symmetric NCS, where a non-trivial topological phase exists only when the triplet pairing is larger than the singlet one \cite{santos_prb_2010}. We find that even though the excitation gap of the superconducting system vanishes at a certain value of $\Delta_t/\Delta_s$ and is non-zero on either side of this axis, the system has precisely the same non-Abelian topological properties on both sides of the gap closing point. The superconducting states on both sides of the gap closing point are characterized by order parameter defects carrying zero-energy non-Abelian Majorana fermion excitations. There is no quantum phase transition, not even of the topological type, even though the excitation gap vanishes as the parameter $\Delta_t/\Delta_s$ is tuned! The only topological phase transition that takes place in this system happens when the chemical potential or the Zeeman splitting are tuned to values such that $V_z^2=\mu^2+\Delta_s^2$. For $V_z$ smaller than the critical value $V_z^c = \sqrt{\mu^2+\Delta_s^2} $ the system is a regular non-centrosymmetric superconductor with a mixture of $s$-wave and $p$-wave pairing amplitude. On the other hand, for $V_z$ greater than the critical value $V_z^c$, the system is a non-Abelian non-centrosymmetric superconductor with order parameter defects such as vortices carrying non-degenerate Majorana modes.

We next look for non-degenerate Majorana solutions in vortex cores in non-centrosymmetric superconductors by explicitly solving the relevant BdG equations.  We find that the same condition on the Zeeman splitting, chemical potential, and (only) the $s$-wave component of the order parameter needs to be satisfied for a unique Majorana solution to exist. As a corollary of this calculation, we find that the existence of the zero-energy Majorana modes in the bulk defects such as vortices is intricately related to the existence of gapless edge modes. The existence condition for the non-Abelian state involves the quantity $(V_z^2 - \mu^2 - \Delta_s^2)$, which is really the Pfaffian of the BdG Hamiltonian at momentum $|\mathbf{k}|=0$. Since the $p$-wave pairing amplitude in the Hamiltonian does not contribute to this Pfaffian, we arrive at the important result that the non-centrosymmetric superconductors are always in the non-Abelian phase as long as $\Delta_s$ satisfies $V_z^2 > \mu^2+\Delta_s^2$, \emph{irrespective of} the value of the $p$-wave component of the order parameter.

    We have derived the important result that the non-Abelian properties of the non-centrosymmetric superconductors are completely insensitive to the magnitude of the $p$-wave piece of the order parameter by seemingly two independent methods: by using a $\mathbb{Z}$ topological invariant and by the direct solutions of the BdG equations which gives us a condition involving the Pfaffian of the BdG Hamiltonian at $k=0$. In the last part of the paper, we make the connection between the topological invariant, which is nothing but the TKNN number, and the Pfaffian of the 1D BdG Hamiltonian for the Majorana mode. The Pfaffian is shown to be related to the first Chern number through the Berry connection. The fact that the superconductivity is intrinsic (\emph{i.e.,} it does not need to be induced by the proximity effect) and the non-Abelian phase can exist for any value of the triplet pairing amplitude makes non-centrosymmetric superconductors a promising candidate for the realization of Majorana bound states and non-Abelian statistics.
\acknowledgements{We thank T. Neupert, M. Sato and S. Fujimoto for correspondence. P.G. wishes to thank Kai Sun for valuable discussions on topological invariants. J.S thanks X.-L. Qi for discussions. P.G. is supported by National Institute of Standards and Technology through Grant Number 70NANB7H6138, Am 001 and through Grant Number N000-14-09-1-1025A by the Office of Naval Research. J.D.S. and S.D.S. are supported by DARPA-QuEST, JQI-NSF-PFC, and LPS-NSA. S.T. acknowledges support from DOE/EPSCoR Grant Number DE-FG02-04ER-46139 and Clemson University start up funds.}


\end{document}